\documentclass[notitlepage,superscriptaddress,showpacs,a4paper,twocolumn,nobalancelastpage,aps,prl,longbibliography]{revtex4-2}
\usepackage[english]{babel}
\RequirePackage[T1]{fontenc}
\RequirePackage{times} 
\usepackage[subpreambles=true]{standalone}

\usepackage{siunitx} 
\usepackage{amsfonts}
\usepackage{amsmath}
\usepackage{amssymb}
\usepackage{graphicx,epstopdf}
\usepackage{appendix}
\usepackage{xspace}
\usepackage{array}
\usepackage[hidelinks]{hyperref}
\usepackage{color}
\usepackage{xcolor}
\usepackage{bm}
\usepackage{verbatim}
\usepackage{tikz}
\usepackage{float}
\usepackage{appendix}
\usepackage{pdfpages}
\usepackage{lipsum}
\allowdisplaybreaks[3]

\makeatletter
\AtBeginDocument{\let\LS@rot\@undefined}
\makeatother
\usetikzlibrary{calc,3d}
\usetikzlibrary{arrows}
\usetikzlibrary{snakes}

\usepackage[normalem]{ulem}

{\par}
\newcounter{mycomment}

\newcommand\rmv{\bgroup\markoverwith {\textcolor{red}{\rule[0.5ex]{2pt}{0.4pt}}}\ULon}

\begin{document}

\title{Frequency modulation on magnons in synthetic dimensions}

\author{Meng Xu}
\affiliation{Department of Physics, Fudan University, Shanghai 200433, China}

\author{Yan Chen}
\email{yanchen99@fudan.edu.cn}
\affiliation{Department of Physics, Fudan University, Shanghai 200433, China}
\affiliation{State Key Laboratory of Surface Physics, Fudan University, Shanghai 200433, China}

\author{Weichao Yu} 
\email{wcyu@fudan.edu.cn}
\affiliation{Institute for Nanoelectronic Devices and Quantum Computing, Fudan University, Shanghai 200433, China}
\affiliation{State Key Laboratory of Surface Physics, Fudan University, Shanghai 200433, China}
\affiliation{Zhangjiang Fudan International Innovation Center, Fudan University, Shanghai 201210, China}

\begin{abstract}
Magnons are promising candidates for next-generation computing architectures, offering the ability to manipulate their amplitude and phase for information encoding. However, the frequency degree of freedom remains largely unexploited due to the complexity of nonlinear process. In this work, we introduce the concept of synthetic frequency dimension into magnonics, treating the eigenfrequency of inherent modes as an additional degree of freedom. This approach enables the effective description of the temporal evolution of a magnon state using an effective tight-binding model, analogous to a charged particle hopping in a modulated lattice. A magnonic ring resonator is investigated as an example, and several intriguing phenomena are predicted, including Bloch oscillations and a leverage effect during unidirectional frequency shifts, all of which are verified through micromagnetic simulations. Notably, our strategy operates in the linear spin-wave regime, excluding the involvement of multi-magnon scattering and high-power generation. This work expands the toolkit for designing magnonic devices based on frequency modulation and paves the way for a new paradigm called magnonics in synthetic dimensions.
\end{abstract}

\maketitle

\textit{Introduction}. Magnons or spin waves are elementary exciations in magnetic systems free from Joule heating, which are promising information carriers for building magnonic circuits \cite{wang_nanoscale_2024} with potential applications for classical \cite{yu_magnetic_2020}, neuromorphic \cite{grollier_neuromorphic_2020} and other unconventional computing architectures \cite{chumak_advances_2022} etc. During past few decades, efforts have been made to explore possible ways to encode information into intrinsic degrees of freedom of magnons. Similar to other wave counterparts such as acoustic waves and electromagnetic waves, there are mainly four routes to encode information in spin waves: (i) Amplitude, where binary information 1(0) is encoded into spin waves with high (low) amplitude and by manipulation of which one can design unidirectional devices and logic gates \cite{lan_spin-wave_2015,chumak_magnon_2014}. (ii) Phase, where the concept of wave interference is harnessed and devices like Mach-Zehnder-type spin wave interferometer \cite{lee_conceptual_2008} and majority gates \cite{klingler_design_2014} are proposed. (iii) polarization, which is a degree of freedom possessed by antiferromagnets and ferrimagnets with two opposite sublattices \cite{lan_antiferromagnetic_2017,han_coherent_2023}. (iv) Frequency, where information is encoded into spectral components of time-varying signals with typical application of magnonic frequency combs \cite{wang_enhancement_2024,xu_magnonic_2023,wang_magnonic_2021,wang_twisted_2022}.

Despite the widespread use of frequency modulation in radio communication benefiting from its great bandwidth efficiency and less susceptibility to interference, there have been few magnonic logic devices utilizing this concept due to lack of strategies to perform frequency modulation on magnons efficiently. Recently, the development of nonlinear magnonics \cite{zheng_tutorial_2023} offers possible ways to manipulate magnonic state in frequency domain. One typical strategy is to realize spectral shift when a magnetic system is driven into nonlinear regime by high-power inputs \cite{li_nutation_2019,wang_nonlinear_2020,lake_interplay_2022}. Another strategy is to trigger the nonlinear scattering process between magnon modes and other modes, such as breathing mode of magnetic skyrmion \cite{wang_magnonic_2021,jin_nonlinear_2023}, cavity photons \cite{wang_bistability_2018,lee_nonlinear_2023} and a pump-induced magnon mode \cite{rao_unveiling_2023,xu_magnonic_2023}. Unfortunately, the involvement of nonlinear process requires that the amplitude of spin waves must be larger than certain threshold, leading to instability and chaotic dynamics \cite{zheng_tutorial_2023}, against the intention for energy efficiency and feasible controllability.

In this Letter, we apply the concept of \textit{synthetic dimension} to perform frequency modulation on magnons in linear regime. Instead of shifting the spectrum itself, the proposed strategy aims to redistribute magnon occupation on the spectrum by using a temporally periodic driving field, in accordance with Floquet engineering \cite{xu_floquet_2020,yang_theory_2023,li_floquet_2023}. The concept of synthetic frequency dimension was first proposed by Yuan et al. \cite{yuan_bloch_2016,yuan_photonic_2016,yuan_synthetic_2018,yuan_synthetic_2021,dutt_single_2020}, which has been successfully applied in photonics \cite{dutt_experimental_2019,qin_spectrum_2018,qin_discrete_2020,chen_tight-binding_2021} and atomic trap systems \cite{oliver_bloch_2023}. The key spirit of synthetic dimension is to treat an intrinsic degree of freedom, e.g., the frequency dimension, on an equal footing with spatial dimension, so that one can manipulate temporal dynamics of excitations such as photons and magnons in the perspective of particle transport in an one-dimensional lattice. 

\begin{figure}[t]
  \centering
  \includegraphics[width=7.6cm]{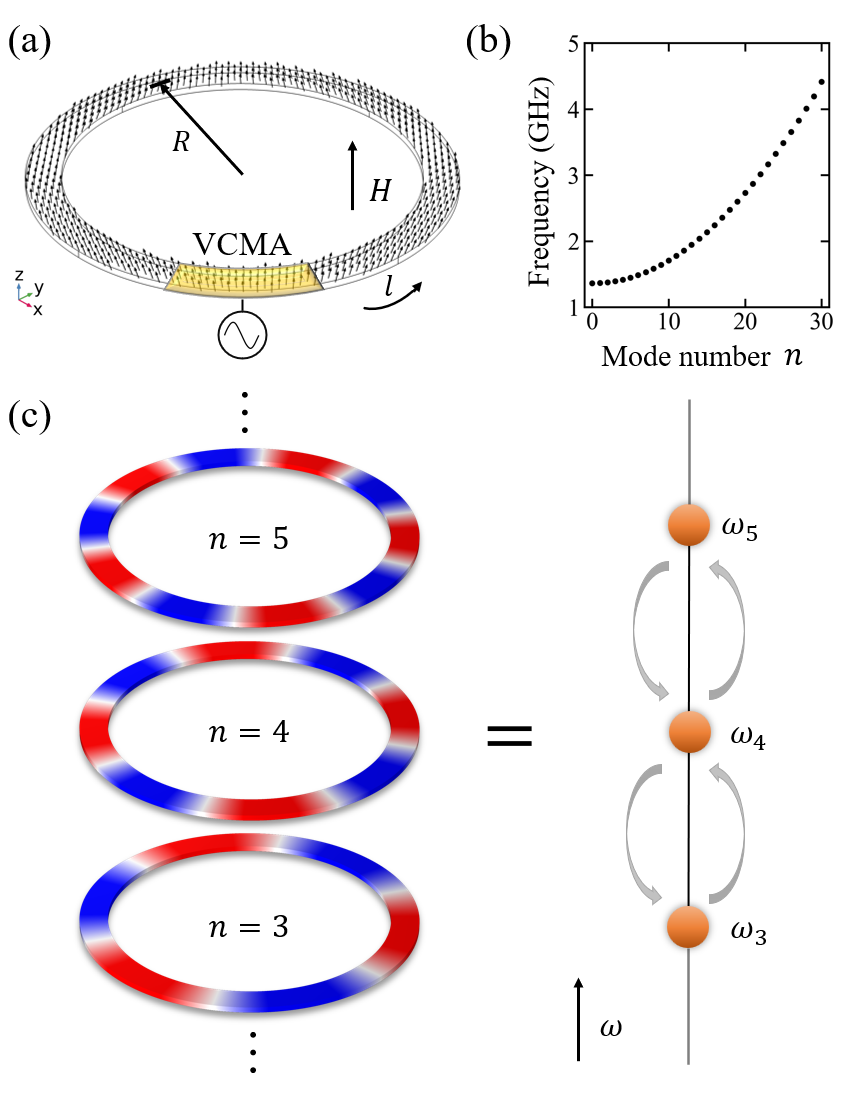}
  \caption{(a) Schematics for a magnonic ring resonator attached by a modulator with alternating voltages. (b) Resonant modes of the magnonic resonator in (a) with unequal frequency spacing. (c) Mode number can be treated as discrete lattices in synthetic frequency dimension so that the evolution of spin-wave excitation is equivalent to a particle hopping in the synthetic lattices.}
  \label{fig:1}
\end{figure}

As depicted in Fig.\ref{fig:1}(a), we consider a two-dimensional magnonic ring resonator with central radius $R$ and width $w$. The magnetizations are uniformly oriented along $\hat{\mathbf{z}}$ by an external field $H$. Such a ring resonator supports unequally spaced resonant modes due to the group velocity dispersion naturally preserved by spin waves, as shown in Fig.\ref{fig:1}(b). Denoting $n$ as mode number for resonant modes with eigenfrequencies $\omega_n$, each mode can be treated as a lattice site in the synthetic frequency dimension, so that the transition of magnonic states in a ring resonator is equivalent to hopping of magnons in a synthetic one-dimensional lattice, as illustrated in Fig.\ref{fig:1}(c). In order to trigger magnon hopping between synthetic lattices, a dynamical modulation is needed which is proposed here to be an alternating voltage applied on the yellow region in Fig.\ref{fig:1}(a), based on the concept of voltage-controlled magnetic anisotropy (VCMA) \cite{maruyama_large_2009,rana_voltage-controlled_2018,rana_towards_2019}, thus inducing a local change of effective field.

\textit{Effective tight-binding model}.  According to the micromagnetic theory \cite{stancil_spin_2009}, the dynamics of the unit magnetization $\mathbf{m}$ is governed by the Landau-Lifshitz-Gilbert (LLG) equation \cite{stancil_spin_2009}
\begin{equation} 
\dot{\mathbf{m}}(\mathbf{r}, \mathrm{t})=-\gamma \mathbf{m}(\mathbf{r}, \mathrm{t}) \times \mathbf{H}_{\text {eff }}+\alpha \mathbf{m}(\mathbf{r}, \mathrm{t}) \times \dot{\mathbf{m}}(\mathbf{r}, \mathrm{t}),
\label{eq:LLG}
\end{equation}
with gyromagnetic ratio $\gamma$ and Gilbert damping coefficient $\alpha$. The effective field $\mathbf{H}_{\text{eff}}=A\nabla^2\mathbf{m}+H\hat{\mathbf{z}}$ includes exchange interaction and external field. Demagnetizing field is not considered here for simplicity, which is valid for high-frequency excitations where exchange interaction dominates. Assuming spin waves are uniformly excited in radial direction and propagate tangentially along arc $l$, Eq.(\ref{eq:LLG}) can be linearized by decomposing the excitation into static and dynamical components as $\mathbf{m}(l,t)=\mathbf{m}_0+\delta\mathbf{m}e^{i(\omega t-kl)}$, and one can obtain the linearized equation of motion for the right-handed precession mode in the absence of Gilbert damping
\begin{equation} 
-\omega m_{+}  = -\gamma H m_{+} - \gamma A k^2 m_{+},
\label{eq:LLGlinearized}
\end{equation}
with $m_+\equiv(\delta m_x+i\delta m_y)e^{i(\omega t-kl)}$. Dispersion of spin-wave eigenstates can be obtained from Eq.(\ref{eq:LLGlinearized}) that $\omega_n=\omega_0+\gamma A k_n^2=\omega_0+\omega^\prime n^2$ with $\omega_0=\gamma H$, $\omega^\prime=4\pi^2\gamma A/L^2$ and $k_n=(2\pi n)/L$ due to the periodic nature of ring structure. It's seen in Fig.\ref{fig:1}(b)
that the magnonic ring resonator supports a set of modes with unequal frequency spacing, which is a distinct feature of magnons other than other excitations with linear dispersions such as photons \cite{kippenberg_microresonator-based_2011}.

Considering a dynamical modulation is applied with strength $H_\text{m}$, angular frequency $\Omega$ and initial phase $\phi$, Eq.(\ref{eq:LLGlinearized}) can be further rearranged as a Schr\"{o}dinger-like equation 

\begin{equation} 
i \dot{m}_{+} =\gamma A \nabla^2 m_{+}-\gamma\left[H+H_\text{m} \cos (\Omega t+\phi)\right] m_{+},
\label{eq:Schrodinger}
\end{equation}
which governs the motion of wave function $m_+$ in a periodic time-varying potential.

\begin{figure*}[htb]
  \centering
  \includegraphics[width=18cm]{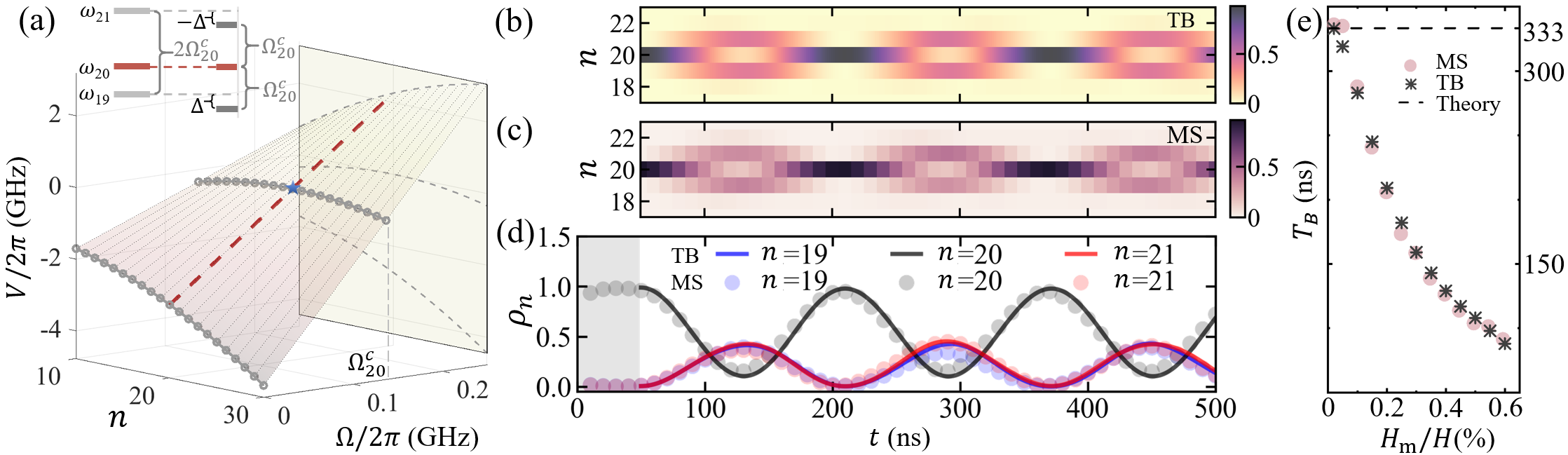}
\caption{(a) Effective onsite potential of magnons evolving on a synthetic frequency lattice under dynamical modulation. The inset illustrates the energy diagram for a magnon state initially placed at $\omega_{20}$ when Bloch oscillation is achieved with the critical driving frequency $\Omega=\Omega_{20}^c$. (b)-(d) Magnonic Bloch oscillation in synthetic frequency dimension, calculated by (b) tight-binding (TB) model and (c) micromagnetic simulation (MS), where a dynamical modulation with strength $H_\text{m}$ and angular frequency $\Omega=\Omega_{20}^c$ is continuously applied after $t=\,$\SI{50}{ns}. The occupation of magnon state is characterized by probability density $\rho_n$ from TB and normalized spin-wave spectrum from MS. (e) Bloch period $T_\text{B}$ versus modulation strength $H_\text{m}$. The dashed line indicates theoretical Bloch period $T_\text{B}=2\pi/|\Delta_{20}|$ for a synthetic lattice with equally spaced levels.}
\label{fig:2} 
\end{figure*}

In general, spin-wave excitation $m_+$ can be expressed as the superposition of all resonant modes with mode amplitude $C_n$, i.e.,
\begin{equation}
m_{+}(l, t)=\sum_n C_n(t) e^{i\left(\omega_n t-k_n l\right)}.
\label{eq:superposition}
\end{equation}

Substituting Eq.(\ref{eq:superposition}) into Eq.(\ref{eq:Schrodinger}) by setting $\phi=0$ and keeping leading-order terms with assumption that $\omega_n-\omega_{n-1}<\Omega<\omega_{n+1}-\omega_n$, one can obtain a coupled-mode equation (see detailed derivation in Supplemental Material (SM) \cite{SM})
\begin{equation}
i \dot{C}_n=g\left[e^{i (\omega_{n+1}-\omega_{n}-\Omega)t} C_{n+1}+e^{-i (\omega_{n}-\omega_{n-1}-\Omega)t} C_{n-1}\right],
\label{eq:coupled-mode}
\end{equation}
with coupling strength $g=-\gamma H_\text{m}/2$ proportional to the modulation strength. We denote the magnon state of the ring resonator as $|\psi(t)\rangle=\sum_n C_n(t) a_n^\dagger|0\rangle$, with $a_n^\dagger$ ($a_n$) the creation (annihilation) operator applied on vaccum state $|0\rangle$. According to Eq.(\ref{eq:coupled-mode}), $|\psi(t)\rangle$ evolves following the Schr\"{o}dinger equation $i\partial_t |\psi(t)\rangle=\mathcal{H}|\psi(t)\rangle$ with the effective Hamiltonian
\begin{equation}
\begin{aligned}
&\mathcal{H}=\\
&\sum_n {g} \left[a_n^{\dagger} a_{n+1} e^{i\left(\omega_{n+1}-\omega_{n}-\Omega\right) t}+a_{n+1}^{\dagger} a_n e^{-i\left(\omega_{n+1}-\omega_{n}-\Omega\right) t}\right].
\label{eq:Hamiltonian0}
\end{aligned}
\end{equation}

We perform gauge transformation on the creation (annihilation) operators $\tilde{a}^\dagger_n(t)=\exp[-i(\omega_n-n\Omega)t]a^\dagger_n(t)$, so that the Hamiltonian Eq.(\ref{eq:Hamiltonian0}) can be gauge-transformed according to $\tilde{\mathcal{H}}=\mathcal{U}(t)^\dagger\mathcal{H}\mathcal{U}(t)-i\mathcal{U}^\dagger(t) \mathrm{d}_t \mathcal{U}(t)$ \cite{eckardt_high-frequency_2015,SM}, which reads as
\begin{equation} 
\tilde{\mathcal{H}}=g \sum_n\left(\tilde{a}_n^{\dagger} \tilde{a}_{n+1}+\tilde{a}_{n+1}^{\dagger} \tilde{a}_n\right)+\sum_n (n \dot \Omega t+n \Omega -\omega_{n}) \tilde{a}_n^{\dagger} \tilde{a}_n,
\label{eq:Hamiltonian1}
\end{equation}
where we assume the frequency of modulation $\Omega(t)$ to be time-dependent without loss of generality. Equation (\ref{eq:Hamiltonian1}) is the effective tight-binding Hamiltonian describing the dynamics of a particle in a synthetic lattice with a time-independent hopping rate $g$ and a site-dependent potential $V_n(t)=n \dot \Omega t+n \Omega -\omega_{n}$.  Other forms of effective Hamiltonian with different physical interpretations can be obtained by choosing other gauges \cite{SM}. 

In the rest of the Letter, the effective tight-binding Hamiltonian Eq.(\ref{eq:Hamiltonian1}) is solved by QuTiP \cite{johansson_qutip_2013} and the original LLG equation Eq.(\ref{eq:LLG}) is solved by Micromagnetics Module based on COMSOL Multiphysics \cite{comsol,zhang_frequency-domain_2023}. The following micromagnetic parameters of Yittrium Iron Garnet (YIG) are considered \cite{lan_spin-wave_2015}: $A=\,$\SI{0.328e-10}{A\,m}, $\gamma=\,$\SI{2.21e5}{Hz/(A/m)}, $\alpha=\,$\SI{1.3e-4}{} and $H=\,$\SI{0.388e5}{A/m} which is equivalent to the intrinsic crystalline anisotropy. The central radius of the ring resonator $R=\,$\SI{575}{nm} with width $w=\,$\SI{50}{nm}. The dynamical modulation is applied to the sector region of the ring resonator with angle of 20 degrees (see SM \cite{SM} for details of numerical methods).

\textit{Bloch oscillation}. Assuming that the frequency of dynamical modulation is temporally invariant, the effective onsite potential can be simplified as $V_n=n \Omega -\omega_{n}$ according to Eq.(\ref{eq:Hamiltonian1}), which is plotted in Fig.\ref{fig:2}(a). It is seen that the dynamical modulation applies a frequency-dependent modification on the effective potential in the synthetic frequency lattice. The gradient of onsite potential $\Delta_n=V_n-V_{n-1}=\Omega-(\omega_n-\omega_{n-1})$ plays as an effective electric field $E_\text{eff}\sim \partial V_n/\partial n$ applied on a charged particle \cite{yuan_synthetic_2021}. When $\Omega/2\pi=\,$\SI{0}{GHz}, the spacing of onsite potential between neighbour sites forms an arithmetic progression due to the quadratic dispersion of exchange spin waves. In the presence of nonzero modulation strength $H_\text{m}$, it is expected that magnon occupation on synthetic lattices may experience complex evolution where analytical prediction is not always possible. However, regular evolution can be achieved at certain specific working point, e.g., \textit{Bloch oscillation} of magnons in synthetic frequency dimension. The conventional realization of Bloch oscillation on charged particles requires an homogeneous electric field \cite{bloch_uber_1929, krieger_time_1985}, or alternatively the presence of Wannier-Stark ladders with equidistant energies \cite{hartmann_dynamics_2004,tartakovskaya_wannier-stark_2020}. In our magnonic system, Wannier-Stack ladders can be approximately achieved when $|\Delta_n|=|\Delta_{n-1}|$, which gives the critical modulation frequency 
\begin{equation} \label{eq:criticalfrequency}
    \Omega_n^c=(\omega_{n+1}-\omega_{n-1})/2.
\end{equation}

For the magnonic state initiallay placed at $\omega_{20}$, the critical modulation frequency is $\Omega_{20}^c/2\pi=\,$\SI{135}{MHz} as labelled in Fig.\ref{fig:2}(a). As depcited in the inset of Fig.\ref{fig:2}(a), the gradient of onsite potential $\Delta_n$ is essentially the offset between original magnonic resonate state and the dressed state in the presence of dynamical modulation. We further perform numerical calculation based on tight-binding (TB) model (Eq.(\ref{eq:Hamiltonian1})) and micromagnetic simulation (MS) (Eq.(\ref{eq:LLG})). As shown in Fig.\ref{fig:2}(b)-(d), a magnonic state at $\omega_{20}$ is initially prepared \cite{SM} and is characterized by probability density $\rho_n=|\langle n | \psi(t)\rangle|^2$ for tight-binding model and normalized spin-wave spectrum extracted from micromagnetic simulation (see SM \cite{SM} for numerical methods). A dynamical modulation with frequency $\Omega_{20}^c$ and strength $H_\text{m}=0.3\%H$ is turned on at $t=\,$\SI{50}{ns}. The magnon state starts its coherent evolution by widening and shrinking with Bloch period $T_\text{B}$, populating an interval from $\omega_{19}$ to $\omega_{21}$. The quantitative agreement shown in Fig.\ref{fig:2}(d) between two methods indicates the validity of the effective tight-binding model. 

The dependence of Bloch period $T_\text{B}$ on the strength of dynamical modulation $H_\text{m}$ is further investigated. According to Eq.(\ref{eq:Hamiltonian1}), the strength of modulation $H_\text{m}$ has no contribution on the onsite potential $V_n$ but plays a crucial role on the hopping rate $g$. With a larger $H_\text{m}$ (associated with larger $g$), the evolving interval from origin will be extended, proportional to $2g/\Delta_n$ \cite{hartmann_dynamics_2004}. However, due to the unequal level spacing of the magnonic system, multiple hopping process will be involved as the magnon state hops to neighbour sites, with effective electric field $\cdots, \Delta_{18},\Delta_{19},\Delta_{21},\Delta_{22} \cdots$, and so on, which are larger than $\Delta_{20}$. Hence it can be expected that a larger $H_\text{m}$ would lead to decreasing Bloch period $T_\text{B}$. The overall Bloch period is estimated using both TB and MS methods and the results are plotted in Fig.\ref{fig:2}(e), which agrees with the expectation. When $H_\text{m}$ approaches to zero, the Bloch period converges to the theoretical value $T=2\pi/|\Delta_{20}|\simeq\,$\SI{333}{ns} (dashed line in Fig.\ref{fig:2}(e)) predicted for a conventional charged particle in a lattice with equally spaced potentials \cite{hartmann_dynamics_2004}. It is worth noting that when the evolving interval is further increased, the assumption that $\omega_n-\omega_{n-1}<\Omega<\omega_{n+1}-\omega_n$ no longer holds, and long-range hopping becomes possible.This would enable the direct hopping of magnon states to farther sites, such as next-nearest neighbor sites. Long-range hopping is inherently accounted for in the micromagnetic simulation but is not captured by the current model described in Eq.(\ref{eq:Hamiltonian1}) (see SM for discussion on long-range hopping process \cite{SM}).

\begin{figure}[h]
  \centering
  \includegraphics[width=8.5cm]{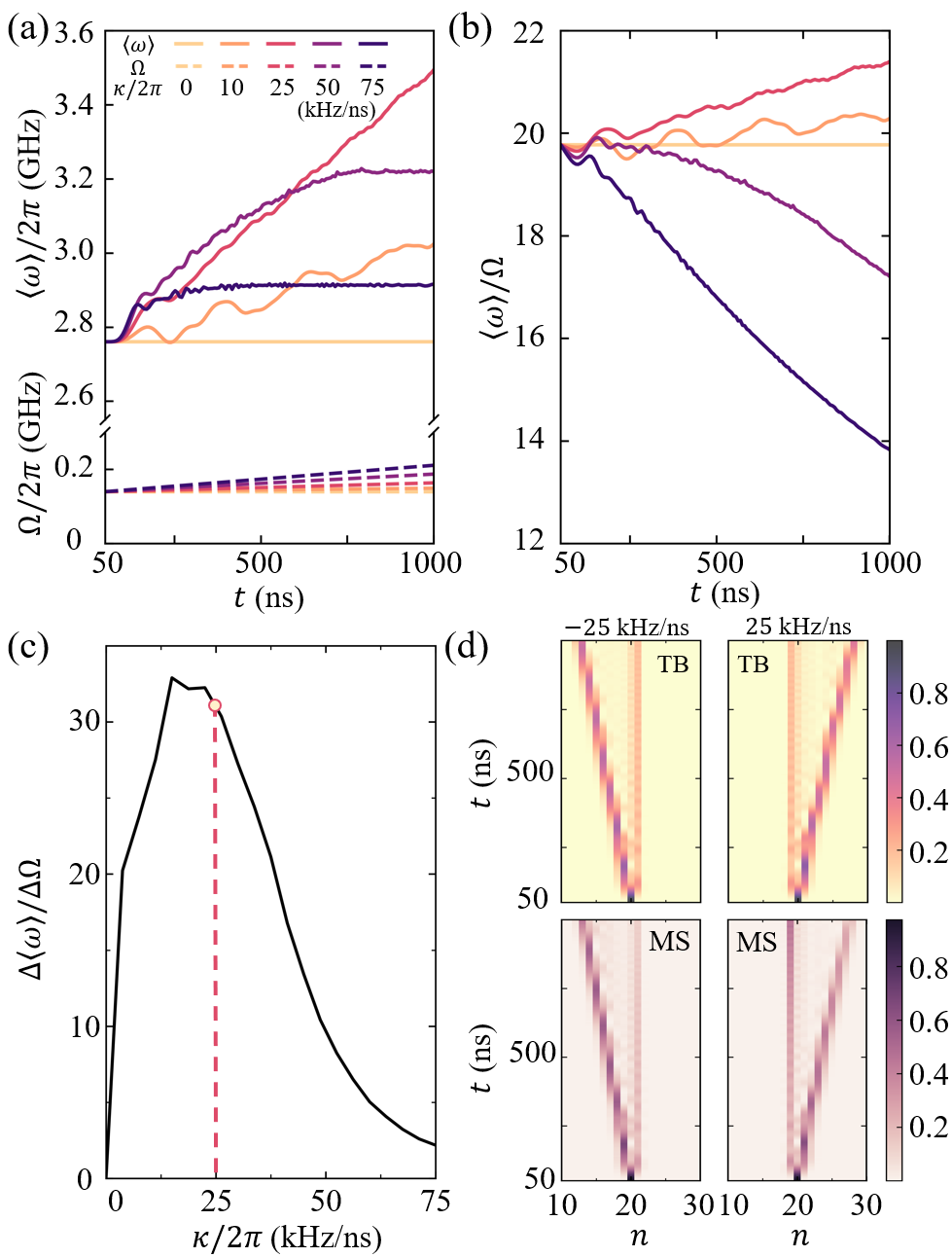}
\caption{ (a) Temporal evolution for center of occupation $\langle\omega\rangle$ started from state $\omega_{20}$ under dynamical modulation with frequency linearly increasing with time $\Omega(t)=\Omega_0+\kappa (t-t_0)$ which is applied after $t_0=\,$\SI{50}{ns}. (b) Frequency ratio $\langle\omega\rangle/\Omega$ versus time. (c) Leverage ratio $\Delta\langle\omega\rangle /\Delta\Omega$ versus changing rate of modulation $\kappa$. (d) Numerical results obtained from TB and MS methods for $\kappa/2\pi=\pm$\SI{25}{kHz/ns}, demonstrating the symmetric frequency shift of magnon state in both directions.
}
\label{fig:3} 
\end{figure}

\textit{Leverage effect for unidirectional frequency shift}. Other than periodic evolution such as Bloch oscillation, unidirectional shift of magnon state can be realized if the dynamical modulation is time-dependent. According to Fig.\ref{fig:2}(a), the effective onsite potential in the absence of dynamical modulation follows an arithmetic sequence, hence we propose that unidirectional frequency shift can be achieved by a dynamical modulation whose frequency increases or decreases linearly with time, i.e., $\Omega(t)=\Omega_0+\kappa t$ with the initial driving frequency $\Omega_0$ and the changing rate $\kappa$. Figure.\ref{fig:3}(a) shows the unidirectional evolution of magnon state starting from $\omega_{20}$, driven by a dynamical modulation with initial frequency $\Omega_0=\Omega_{20}^c$ and changing rate ranging from \SI{0}{kHz/ns} to \SI{75}{kHz/ns}. It is observed that the magnon state, characterized by center of occupation defined as $\langle\omega\rangle=\omega_0+ \langle n \rangle^2\omega^\prime$ with $\langle n \rangle=\sum_n n\rho_n$, is driven unidirectionally from the initial state $\omega_{20}$ towards $\omega_{28}$ within the time interval of \SI{950}{ns} for $\kappa/2\pi=\,$\SI{25}{kHz/ns}. For larger or smaller $\kappa$, frequency shift can be partially achieved, but cannot sustain under a continuous driving. 

It is important to note that there is an order-of-magnitude disparity between the driving frequency $\Omega$ and the resonant frequency $\langle\omega\rangle$, as depicted in Fig.\ref{fig:3}(b), indicating an overall amplification of frequency shift, or a leverage effect,  enabling the realization of a substantial frequency shift on a target by employing a source with a much smaller frequency change. We further define the leverage ratio $\Delta\langle\omega\rangle/\Delta\Omega$ as the ratio of frequency change during the time interval from $t_1=\,$\SI{50}{ns} to $t_2=\,$\SI{1000}{ns} with $\Delta\langle\omega\rangle=\langle\omega\rangle_{t_1}-\langle\omega\rangle_{t_2}$ and $\Delta\Omega=\Omega_{t_1}-\Omega_{t_2}$. The leverage ratio reaches maximum around $\kappa/2\pi=\,$\SI{20}{kHz/ns}, comparable to the average hopping velocity of a particle hopping between neighbour sites $(\Omega_{n}^c-\Omega_{n-1}^c)/2T_\text{B}$. The numerical results for $\kappa/2\pi=$\SI{25}{kHz/ns} produced by both TB and MS methods are shown in Fig.\ref{fig:3}(d), where a symmetric frequency shift in a reversed direction is achieved by letting $\kappa/2\pi=\,$\SI{-25}{kHz/ns}. Except for the state that is driven unidirectionally, there exists another state that remains almost unchanged around the initial state $\omega_{20}$ as a consequence of the inhomogeneous onsite potential, which can be eliminated by preparing the initial magnon state with a finite momentum in the synthetic frequency dimension (see SM \cite{SM} for numerical methods). This study investigates a ring resonator with a geometry on the order of hundred nanometers, which leads to an optimized changing rate around \SI{}{kHz/ns}, too high for realistic application. We highlight that the required changing rate can be further reduced by employing ring resonators with larger geometry, where the spacing between energy levels would be decreased, and the leverage ratio can be further amplified.

\textit{Discussion}. One feature of spin waves is the group velocity dispersion which is naturally present even when dipolar effect is excluded, bringing a vague boundary in the synthetic frequency dimension \cite{shan_one-way_2020} and leading to nonreciprocal evolution of magnon state. Another distinct feature of magnetic systems is the rich dynamics of magnetic textures, such as magnetic vortices \cite{wintz_magnetic_2016} and magnetic skyrmions \cite{schutte_magnon-skyrmion_2014,zhang_frequency-domain_2023}, which supports multiple inherent modes below magnon continuum. It is expected to explore new way to manipulate excitation of either single magnetic texture or quasicrystals such as skyrmion lattices in a hybrid space including realistic spatial dimension and synthetic frequency dimension. Synthetic pseudospin dimension \cite{dutt_single_2020} can be straightforwardly constructed for antiferromagneitc or ferrimagnetic systems where the polarization degree of freedom can be harnessed \cite{lan_antiferromagnetic_2017,yu_magnetic_2021}. Another intriguing topic is the effective gauge field \cite{bell_spectral_2017,fang_realizing_2012} that magnons hop between spatially separated lattices as well as nonreciprocal hopping induced by intrinsic interactions such as Dzyaloshinskii-Moriya interaction \cite{lan_spin-wave_2015}. It needs to be strengthened that the dynamical manipulation is only required during the manipulation of magnon state. When the dynamical modulation is turned off, the effective hopping rate $g$ instantly goes to zero and the magnon state stops to evolve, i.e., stay at where it is rather than returning back to the initial state. We anticipate that our work will stimulate further research on the design of magnonic logic gates and other information processing units based on frequency modulation \cite{maram_frequency-domain_2020,odintsov_lateral_2024}, where binary information is encoded into different magnon occupation in the spectrum rather than in amplitude or phase.

\textit{Conclusion}. We explore the manipulation of magnon states in a ring resonator within the frequency domain by leveraging the concept of synthetic frequency dimensions. By deriving an effective tight-binding model from the original micromagnetic model, we predict the evolution of magnon states, including Bloch oscillations and the leverage effect during unidirectional frequency shifts, which are verified via micromagnetic simulations. Notably, all of these manipulations are valid within the linear regime without the requirement of high-threshold power, which renders them promising for the design of energy-efficient and controllable magnonic devices. Our work potentially opens up a new avenue in magnonics, namely, magnonics in synthetic dimensions.

\section{Acknowledgements}

This work was supported by National Natural Science Foundation of China (Grant No. 12204107 and 12274086), National Key Research Program of China (Grant No. 2022YFA1403300 and 2022YFA1404204), Shanghai Science and Technology Committee (Grant No. 21JC1406200) and Shanghai Pujiang Program (Grant No. 21PJ1401500). The authors thank Luqi Yuan for fruitful discussions.

\appendix

\begin{widetext}

\section{Derivation of the effective tight-binding model}

\subsection{Coupled-mode equation}

We start from the Schr\"{o}dinger-like equation (Eq.(3) in the main text) that 
\begin{equation} 
i \dot{m}_{+} =\gamma A \nabla^2 m_{+}-\gamma\left[H+H_\text{m} \cos (\Omega t+\phi)\right] m_{+},
\label{eq:S1}
\end{equation}
and treat the ``wave function'' $m_+$ as superposition of occupation on each synthetic lattice labeled by $n$
\begin{equation}
m_{+}(l, t)=\sum_n C_n(t) e^{i\left(\omega_n t-k_n l\right)}.
\label{eq:S2}
\end{equation}

Plugging Eq.(\ref{eq:S2}) into Eq.(\ref{eq:S1}) and cancelling the term $e^{i k_n l}$ on both sides, we obtain
\begin{equation}
i \sum_n \left(  \dot{C}_n(t) e^{i\omega_n t} + i \omega_{n} C_n(t) e^{i\omega_n t}\right)=\left(-\gamma A k_n^2-\gamma H\right)\sum_n C_n(t) e^{i\omega_n t}-\gamma H_\text{m} \frac{e^{i\Omega t}+e^{-i\Omega t }}{2} \sum_n C_n(t) e^{i\omega_n t},
\label{eq:S3}
\end{equation}
where we assume $\phi=0$ and adopt the identity relation $2\cos(\Omega t)=e^{i\Omega t}+e^{-i\Omega t }$.

According to the spin-wave dispersion $\omega_n-\gamma A k_n^2-\gamma H=0$ and the denotation $g=-\gamma H_\text{m} /2$, Eq.(\ref{eq:S3}) can be further simplified as
\begin{equation}
i \sum_n  \dot{C}_n(t) e^{i\omega_n t} =
 g \left[ \sum_n C_{n}(t) e^{i\left(\omega_{n}+\Omega \right) t} + \sum_n C_{n}(t) e^{i\left(\omega_{n} -\Omega \right)t}\right].
 \label{eq:S4}
\end{equation}

We focus on the dynamics of $m$-th mode, and Eq.(\ref{eq:S4}) can be expanded as
\begin{equation}
\begin{aligned}
i\Big[\cdots+\dot{C}_{m-1}(t)&e^{i\omega_{m-1}t}+\dot{C}_{m}(t)e^{i\omega_{m}t}+\dot{C}_{m+1}(t)e^{i\omega_{m+1}t}+\cdots\Big]=\qquad\qquad\qquad\qquad\\
&g\left[\cdots+C_{m-1}(t)e^{i(\omega_{m-1}+\Omega)t}+C_{m}(t)e^{i(\omega_{m}+\Omega)t}+C_{m+1}(t)e^{i(\omega_{m+1}+\Omega)t}\cdots\right]+\\
&g\left[\cdots+C_{m-1}(t)e^{i(\omega_{m-1}-\Omega)t}+C_{m}(t)e^{i(\omega_{m}-\Omega)t}+C_{m+1}(t)e^{i(\omega_{m+1}-\Omega)t}\cdots\right].
 \label{eq:S5}
\end{aligned}
\end{equation}

Since the $m$-th mode varies with the angular frequency of $\omega_m$, we eliminate the terms on the left hand side which vary faster or slower than $e^{i\omega_m t}$, and Eq.(\ref{eq:S5}) becomes
\begin{equation}
\begin{aligned}
i \dot{C}_m(t)= g \Big[ \cdots  +\overbrace{C_{m-1}(t) e^{i\left(\omega_{m-1}-\omega_{m}+\Omega \right)t}}^{\text{leading-order term}}+C_{m}(t) e^{i \Omega t}  +C_{m+1}(t) e^{i\left(\omega_{m+1}-\omega_{m}+\Omega \right)t}+ \cdots\Big]+ \\
g\Big[ \cdots  +\underbrace{C_{m+1}(t) e^{i\left(\omega_{m+1}-\omega_{m}-\Omega \right)t}}_{\text{leading-order term}}+C_{m}(t) e^{-i \Omega t}  +C_{m-1}(t) e^{i\left(\omega_{m-1}-\omega_{m}-\Omega \right)t}+ \cdots\Big].
\label{eq:S6}
\end{aligned}
\end{equation}

We assume that the driving frequency of modulation $\Omega$ is close to the energy spacing between neighbour modes around mode $m$, i.e., $\omega_m-\omega_{m-1}<\Omega<\omega_{m+1}-\omega_m$, so that we can keep the slowest terms on the right hand side of Eq.(\ref{eq:S6}). Substituting the label from a specific mode $m$ to arbitrary modes $n$, we can obtain the coupled-mode equation (Eq.(5) in the main text)
\begin{equation}
i \dot{C}_n=g\left[e^{i (\omega_{n+1}-\omega_{n}-\Omega)t} C_{n+1}+e^{-i (\omega_{n}-\omega_{n-1}-\Omega)t} C_{n-1}\right].
\label{eq:S7}
\end{equation}

\subsection{Effective Hamiltonian 1}

The coupled-mode equation Eq.(\ref{eq:S7}) describes the dynamics of a particle characterized by the state $|\psi(t)\rangle=\sum_n C_n(t) a_n^\dagger|0\rangle$ governed by Schr\"{o}dinger equation $i\partial_t |\psi(t)\rangle=\mathcal{H}|\psi(t)\rangle$, with $a_n^\dagger$ ($a_n$) the creation (annihilation) operator applied on vacuum state $|0\rangle$. The effective Hamiltonian reads
\begin{equation}
\mathcal{H}=\sum_n {g} \left(a_n^{\dagger} a_{n+1} e^{i\left(\omega_{n+1}-\omega_{n}-\Omega\right) t}+a_{n+1}^{\dagger} a_n e^{-i\left(\omega_{n+1}-\omega_{n}-\Omega\right) t}\right).
\label{eq:S8}
\end{equation}

We choose the operator ${\mathcal{U}_1}^{\dagger}(t)=\mathrm{diag}[\cdots, e^{i (\omega_{n-1} - (n-1) \Omega)t}, e^{i (\omega_{n} - n \Omega)t}, e^{i (\omega_{n+1} - (n+1) \Omega)t}, \cdots]$ and perform gauge transformation, so that $\tilde{a}^\dagger_n(t)=e^{-i(\omega_n-n\Omega)t}a^\dagger_n(t)$ and $\tilde{a}_n(t)=e^{i(\omega_n-n\Omega)t}a_n(t)$. The effective Hamiltonian (Eq.(7) in the main text) is obtained after gauge transformation $\tilde{\mathcal{H}}=\mathcal{U}_1(t)^\dagger\mathcal{H}\mathcal{U}_1(t)-i\mathcal{U}_1^\dagger(t) \mathrm{d}_t \mathcal{U}_1(t)$ \cite{eckardt_high-frequency_2015}, which is derived as
\begin{equation}
\begin{aligned}
\tilde{\mathcal{H}}  =& \sum_n  g \left[\tilde{a}_n^{\dagger} e^{i (\omega_{n} - n \Omega) t} \tilde{a}_{n+1}^{} e^{-i (\omega_{n+1} - (n+1) \Omega) t} e^{i\left(\omega_{n+1}-\omega_{n}-\Omega\right) t} \right]\\
&+\sum_n  g \left[\tilde{a}_{n+1}^{\dagger} e^{i (\omega_{n+1} - (n+1) \Omega) t} \tilde{a}_{n}^{} e^{-i (\omega_{n} - n \Omega) t} e^{-i\left(\omega_{n+1}-\omega_{n}-\Omega\right) t}\right]\\  
&+ \sum_n\left(-i \frac{\mathrm{d}}{\mathrm{d} t}{(-i (\omega_{n} - n \Omega) t)}\right) \tilde{a}_n^{\dagger} e^{i (\omega_{n} - n \Omega) t} \tilde{a}_n e^{-i (\omega_{n} - n \Omega) t} \\
 = & \sum_n g \left(\tilde{a}_n^{\dagger} \tilde{a}_{n+1}+\tilde{a}_{n+1}^{\dagger} \tilde{a}_n\right) +\sum_n (n \dot \Omega t+n \Omega -\omega_{n}) \tilde{a}_n^{\dagger} \tilde{a}_n. \quad\text{(effective Hamiltonian 1)}
\label{eq:S9}
\end{aligned}
\end{equation}

\subsection{Effective Hamiltonian 2} 

We start from Eq.(\ref{eq:S6}) where have kept only leading-order terms during the derivation above. In this section, we keep more terms with wavy line as shown in Eq.(\ref{eq:S10}) and we can obtain the corresponding coupled-mode equation
\begin{equation}
\begin{aligned}
i \dot{C}_m(t)=& g \Big[ \cdots  +\uwave{C_{m-1}(t) e^{i\left(\omega_{m-1}-\omega_{m}+\Omega \right)t}}+\uwave{C_{m}(t) e^{i \Omega t}}  +\uwave{C_{m+1}(t) e^{i\left(\omega_{m+1}-\omega_{m}+\Omega \right)t}}+ \cdots\Big]+ \\
&g\Big[ \cdots  +\uwave{C_{m+1}(t) e^{i\left(\omega_{m+1}-\omega_{m}-\Omega \right)t}}+\uwave{C_{m}(t) e^{-i \Omega t}}  +\uwave{C_{m-1}(t) e^{i\left(\omega_{m-1}-\omega_{m}-\Omega \right)t}}+ \cdots\Big]\\
=&g (e^{i \Omega t}+e^{-i \Omega t}) \left[e^{i (\omega_{m+1}-\omega_{m})t} C_{m+1}+e^{-i (\omega_{m}-\omega_{m-1})t} C_{m-1}+C_{m}\right],\\
=&2g \mathrm{cos}(\Omega t)  \left[e^{i (\omega_{m+1}-\omega_{m})t} C_{m+1}+e^{-i (\omega_{m}-\omega_{m-1})t} C_{m-1}+C_{m}\right].
\label{eq:S10}
\end{aligned}
\end{equation}

Following the same notation as in Eq.(\ref{eq:S8}), we obtain the effective Hamiltonian according to the coupled-mode equation Eq.(\ref{eq:S10})
\begin{equation}
\begin{aligned}
\mathcal{H}^\prime&=\sum_n 2g\cos (\Omega t) a_n^{\dagger} a_n+\sum_n 2g \cos (\Omega t) \left[a_n^{\dagger} a_{n+1} e^{i\left(\omega_{n+1}-\omega_{n}\right) t}+a_{n+1}^{\dagger} a_n e^{-i\left(\omega_{n+1}-\omega_{n}\right) t}\right]\\
&\sim \sum_n 2g \cos (\Omega t) \left[a_n^{\dagger} a_{n+1} e^{i\left(\omega_{n+1}-\omega_{n}\right) t}+a_{n+1}^{\dagger} a_n e^{-i\left(\omega_{n+1}-\omega_{n}\right) t}\right],
\label{eq:S11}
\end{aligned}
\end{equation}
where the first term $\sum_n 2g\cos (\Omega t) a_n^{\dagger} a_n$ is eliminated because of its independence on lattice site $n$ which will not contribute to the dynamics. By choosing the operator ${\mathcal{U}_2}^{\dagger}(t)=\mathrm{diag}[\cdots, e^{i \omega_{n-1} t}, e^{i \omega_n t}, e^{i \omega_{n+1} t}, \cdots]$, we transform the creation and annihilation operators as $\breve{a}^\dagger_n = e^{-i \omega_n t} a^\dagger_n$ and $\breve{a}_n = e^{i \omega_n t} a_n$, and we can obtain the gauge-transformed Hamiltonian according to $\breve{\mathcal{H}}=\mathcal{U}_2(t)^\dagger\mathcal{H}^\prime\mathcal{U}_2(t)-i\mathcal{U}_2^\dagger(t) \mathrm{d}_t \mathcal{U}_2(t)$, i.e.,

\begin{equation}
\begin{aligned}
\breve{\mathcal{H}} =&\sum_n  2g \cos (\Omega t)\left[ \breve{a}_n^{\dagger} e^{i \omega_n t} \breve{a}_{n+1} e^{-i \omega_{n+1} t} e^{i\left(\omega_{n+1}-\omega_{n}\right) t}+\breve{a}_{n+1}^{\dagger} e^{i \omega_{n+1} t} \breve{a}_n e^{-i \omega_n t}e^{-i\left(\omega_{n+1}-\omega_{n}\right) t}\right]\\
&+\sum_n\left(-i \frac{\mathrm{d}}{\mathrm{d} t}{(-i \omega_n t)}\right) \breve{a}_n^{\dagger} e^{-i \omega_n t} \breve{a}_n e^{i \omega_n t} \\
=&\sum_n 2g \cos (\Omega t)\left(\breve{a}_n^{\dagger} \breve{a}_{n+1}+\breve{a}_{n+1}^{\dagger} \breve{a}_n\right)  - \sum_n \omega_n \breve{a}_n^{\dagger} \breve{a}_n. \quad\text{(effective Hamiltonian 2)}
\label{eq:S12}
\end{aligned}
\end{equation}

Similar to the physical interpretation of Eq.(\ref{eq:S9}), Eq.(\ref{eq:S12}) describes the dynamics of a particle with a time-dependent hopping rate and a time-independent on-site potential. Note that $\omega_n=\omega_0+\omega^\prime n^2$ with $\omega_0=\gamma H$ and $\omega^\prime=4\pi^2\gamma A/L^2$ which increases quadratically with site number $n$. The negative sign in the on-site potential terms rises from the conventional choice of basis $e^{i\omega t}$ when performing Fourier transformation.

\clearpage
\section{Comparison of results calculated from different Hamiltonian}

We have considered different approximations and derived the Hamiltonian $\mathcal{H}$ (Eq.(\ref{eq:S8}) and Eq.(6) in the main text), $\tilde{\mathcal{H}}$ (Eq.(\ref{eq:S9}) and Eq.(7) in the main text) and ${\mathcal{H}^\prime}$ Eq.(\ref{eq:S11}). Two phenomena predicted in the main text, i.e., Bloch oscillation and unidirectional frequency shift, are calculated and the results are compared as shown below. The comparison suggests that the choice of gauge does not influence the validity of the model.

\begin{figure}[h]
  \centering
  \includegraphics[width=17cm]{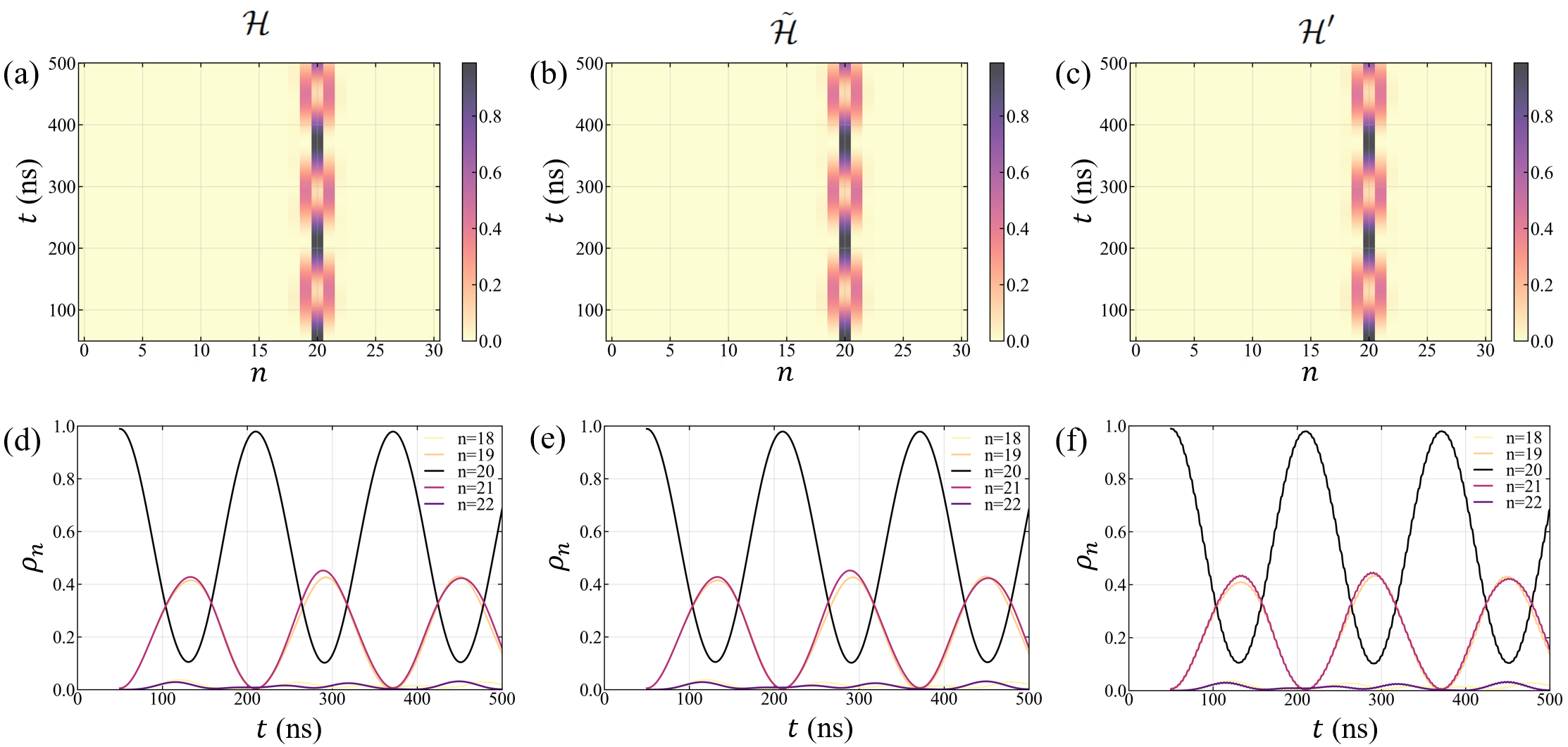}
  \caption{Bloch oscillation calculated from (a)(d) $\mathcal{H}$, (b)(e) $\tilde{\mathcal{H}}$ and (c)(f) ${\mathcal{H}^\prime}$. Parameters are chosen as $H_\text{m}/H=0.3\%$ and $\Omega=(\omega_{21}-\omega_{19})/2$, same as Fig.2(b)-(d) in the main text.}
  \label{fig:S1}
\end{figure}

\begin{figure}[h]
  \centering
  \includegraphics[width=17cm]{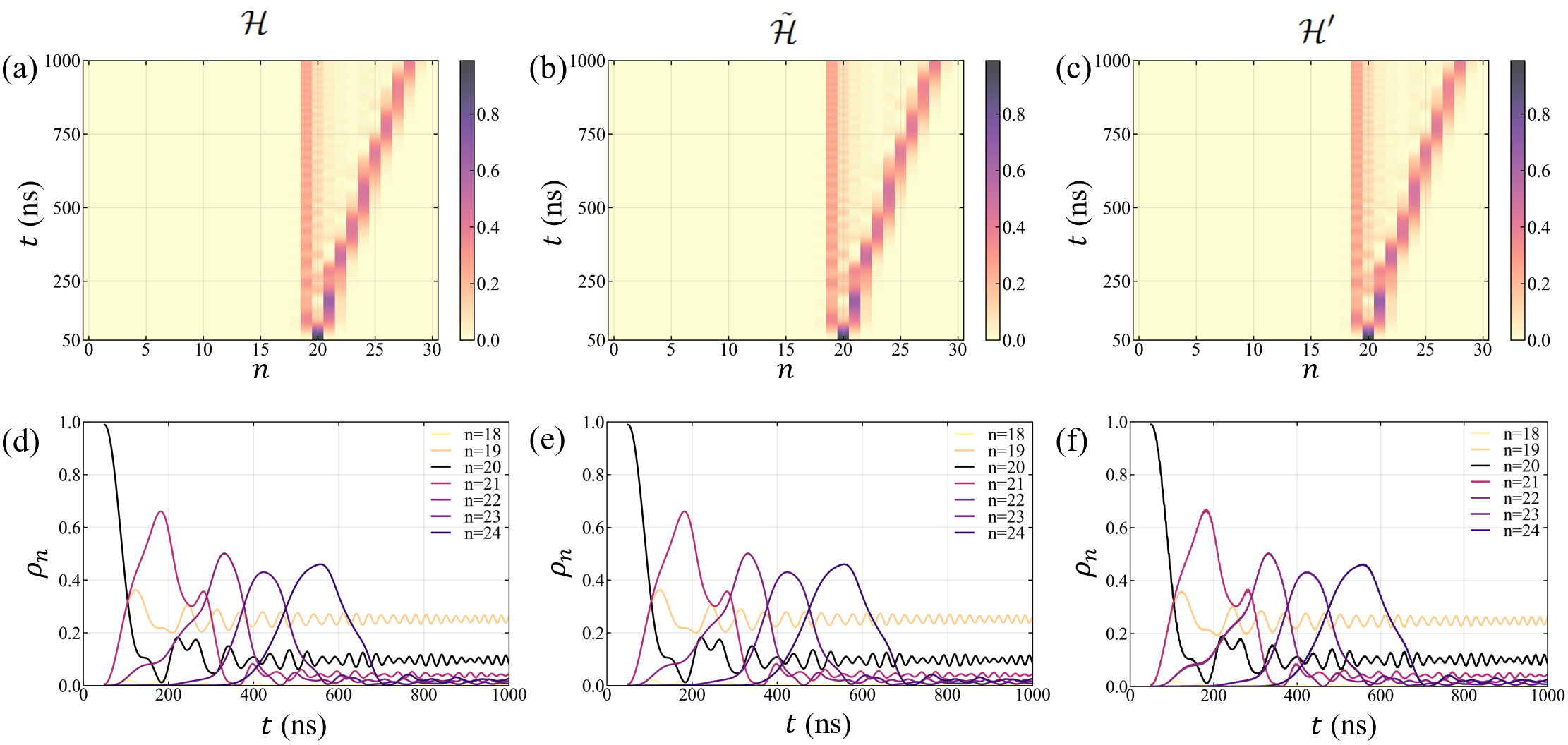}
  \caption{Unidirectional frequency shift calculated from (a)(d) $\mathcal{H}$, (b)(e) $\tilde{\mathcal{H}}$ and (c)(f) ${\mathcal{H}^\prime}$. Parameters are chosen as $H_\text{m}/H=0.3\%$, $\Omega_0=\omega_{20}-\omega_{19}$ and $\kappa=\,$\SI{25}{kHz/ns}, same as Fig.3(d) in the main text.}
  \label{fig:S2}
\end{figure}

\newpage
\section{Coupling between long-range modes}

During the derivation of effective Hamiltonian 1 and 2 (Eq.(\ref{eq:S9}) and Eq.(\ref{eq:S12})), we assume that the driving frequency is close to the spacing between neighbor modes, i.e., $\Omega\sim\omega_{n+1}-\omega_{n}$. When $\omega$ goes larger, the approximation above is no longer valid and one needs to keep more terms in Eq.(\ref{eq:S6}) to characterize the hopping process which skips multiple sites. For example, when $\Omega\sim\omega_{n+2}-\omega_{n}$, the effective Hamiltonian including next-nearest neighbor coupling terms reads
\begin{equation}
\begin{aligned}
\mathcal{H}^{\prime\prime}=&\sum_{n} {g} \left[ a_n^{\dagger} a_{n+1} e^{i\left(\omega_{n+1}-\omega_{n}-\Omega\right) t}+a_{n+1}^{\dagger} a_n e^{-i\left(\omega_{n+1}-\omega_{n}-\Omega\right) t}\right.\\
&\left.+a_n^{\dagger} a_{n+2} e^{i\left(\omega_{n+2}-\omega_{n}-\Omega\right) t}+a_{n+2}^{\dagger} a_n e^{-i\left(\omega_{n+2}-\omega_{n}-\Omega\right) t}\right].
\label{eq:S13}
\end{aligned}
\end{equation}

Figure.\ref{fig:S3} shows the numerical results for Bloch oscillation when the next-nearest coupling is included. It is evidenced that the inclusion of next-nearest coupling does not affect the key phenomena (Fig.2 in the main text) as long as the driving frequency is close to the potential difference between neighbour sites. On the other hand, long-range hopping is possible when the driving frequency is matched and the modulation strength is strong enough.

\begin{figure}[h]
  \centering
  \includegraphics[width=15cm]{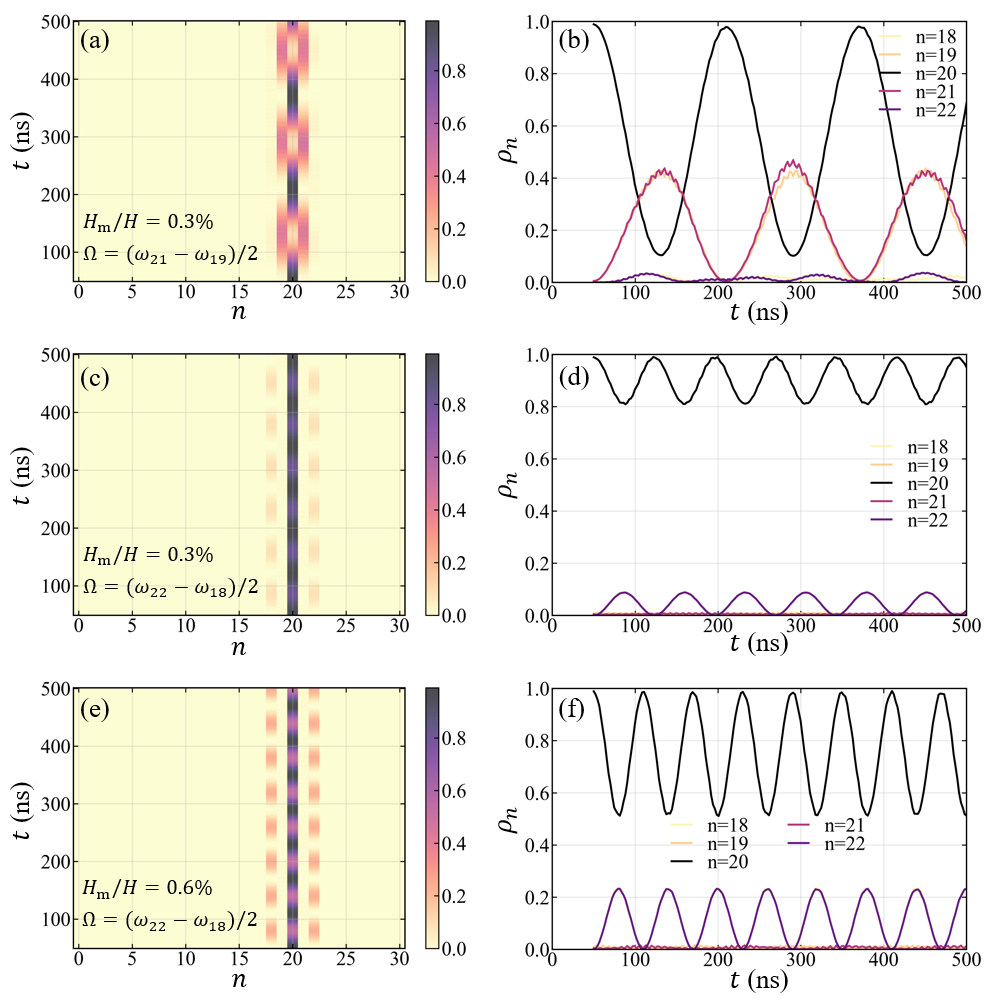}
  \caption{Bloch oscillation calculated from Hamiltonian $\mathcal{H}^{\prime\prime}$ including next-nearest coupling. (a-b) Frequency of dynamical modulation $\Omega=(\omega_{21}-\omega_{19})/2$ and modulation strength $H_\text{m}/H=0.3\%$, working in the same condition as Fig.2 in the main text. (c-d) Frequency of dynamical modulation $\Omega=(\omega_{22}-\omega_{18})/2$ and modulation strength $H_\text{m}/H=0.3\%$. (e-f) Frequency of dynamical modulation  $\Omega=(\omega_{22}-\omega_{18})/2$ and an enlarged modulation strength $H_\text{m}/H=0.6\%$.}
  \label{fig:S3}
\end{figure}

\newpage

\section{Numerical methods}

The effective tight-binding Hamiltonian is solved by QuTiP \cite{johansson_qutip_2013}. The initial state is set as a Gaussian function in the synthetic frequency dimension, i.e., $\langle n | \psi_0\rangle = \frac{1} {\sqrt{\pi} N \sigma} \exp[-\frac{(n-n_0)^2}{(N \sigma)^2}+i k_0 n]$, with the center of initial distribution $n_0=20$, total lattice number $N=30$, the standard deviation $\sigma=0.01$ and the initial momentum $k_0=0$. As pointed out in the main text, a non-zero initial momentum may suppress the unchanged mode. Nevertheless, an unresolved issue pertains to the methodology required for the generation of a magnon state exhibiting initial momentum within the synthetic frequency dimension, whether in the context of experimental realization or micromagnetic simulations.

The micromagnetic simulations are performed using the Micromagnetics Module (Time Domain, V2.12) based on COMSOL Multiphysics \cite{comsol,zhang_frequency-domain_2023}. A magnon state centered at $\omega_{20}$ is injected into the ring resonator by a transverse field pulse with expression $h_0\sin(\omega_{20} t)\exp[-(x/\Delta w)^2]\exp{-[(t-5/f_{20})/(50/f_{20})]^2}$, where $h_0=\,$\SI{1000}{A/m}, $\Delta w=\,$\SI{20}{nm} and $\omega_{20}=2\pi f_{20}$. The dynamical modulation is turned on after $t=\,$\SI{50}{ns} and is applied to the sector region of the ring resonator with angle of 20 degrees, so that there is an amplifying factor  $360^\circ/20^\circ=18$ multiplied on the modulation strength $H_\text{m}$ for the micromagnetic simulations. Figure.\ref{fig:S4}(a) shows the temporal evolution and spatial distribution along arc of unit magnetization component $m_x$. Short-time Fourier transformation is performed on the raw data with 200 windows and overlap factor 0.7, which produces the spectrum shown in Fig. 2(c-d) in the main text.

\begin{figure}[htb]
  \centering
  \includegraphics[width=17cm]{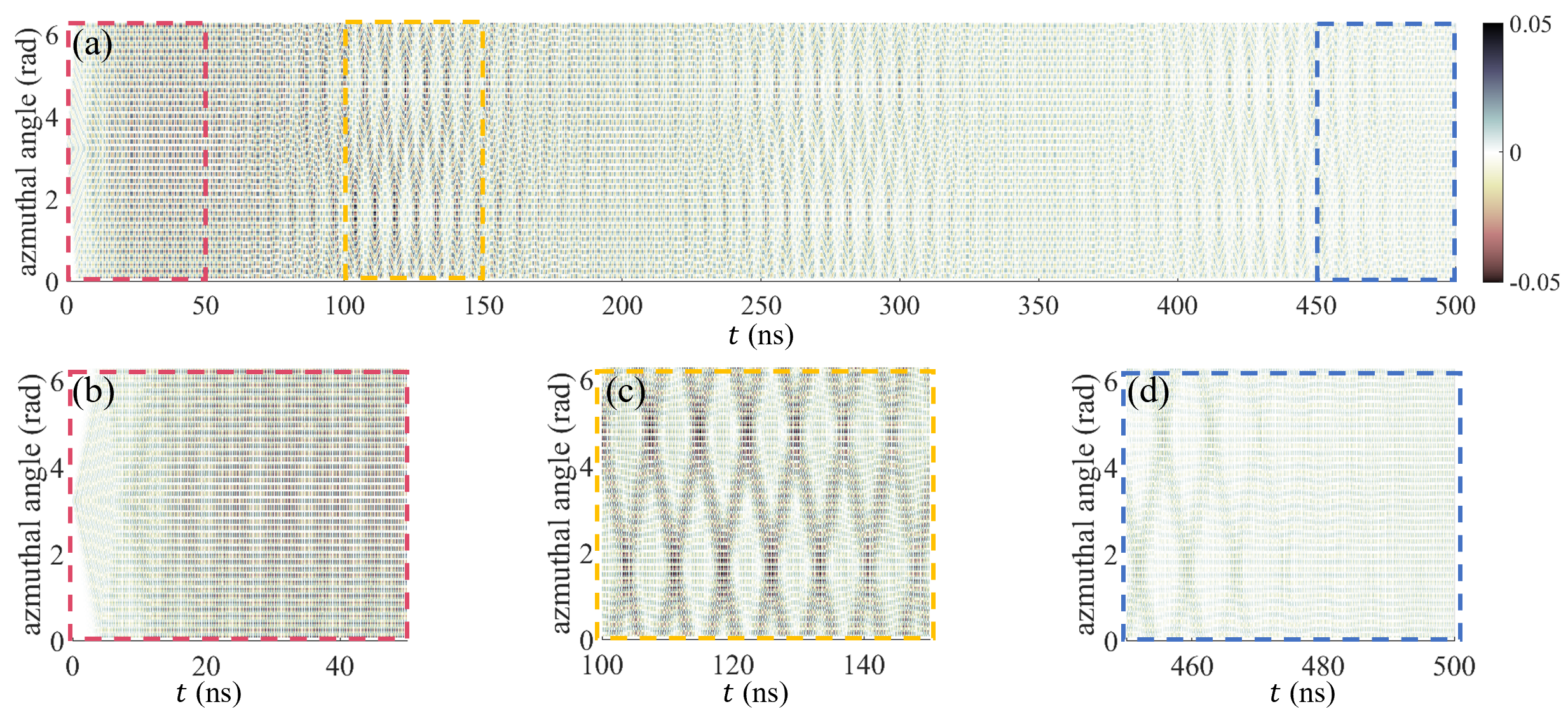}
  \caption{(a) Temporal evolution and spatial distribution of unit magnetization component $m_x$ during the Bloch oscillation scenario (Fig.2 in the main text) obtained from micromagnetic simulation. (b-d) Zoomed-in fragment for different time periods.}
  \label{fig:S4}
\end{figure}

 \end{widetext}



\end{document}